# Pixel-wise Modulated Dice Loss for Medical Image Segmentation

Seyed Mohsen Hosseini [1]

## Abstract

*Class imbalance and the difficulty imbalance are the two types of data imbalance that affect the performance of neural networks in medical segmentation tasks. In class imbalance the loss is dominated by the majority classes and in difficulty imbalance the loss is dominated by easy to classify pixels. This leads to an ineffective training. Dice loss, which is based on a geometrical metric, is very effective in addressing the class imbalance compared to the cross entropy (CE) loss, which is adopted directly from classification tasks. To address the difficulty imbalance, the common approach is employing a re-weighted CE loss or a modified Dice loss to focus the training on difficult to classify areas. The existing modification methods are computationally costly and with limited success. In this study we propose a simple modification to the Dice loss with minimal computational cost. With a pixel level modulating term, we take advantage of the effectiveness of Dice loss in handling the class imbalance to also handle the difficulty imbalance. Results on three commonly used medical segmentation tasks show that the proposed Pixel-wise Modulated Dice loss (PM Dice loss) outperforms other methods, which are designed to tackle the difficulty imbalance problem.*

## 1. Introduction

Class imbalance in a dataset occurs when some classes have significantly fewer examples compared to other classes. This can negatively affect the performance of neural networks in classification and segmentation tasks [20, 17, 28]. With class imbalance the loss function would be biased towards the majority classes, leading to an ineffective training and poor performance on minority classes. This problem is especially severe in medical segmentation tasks where usually the task involves segmenting a small object in a large background [1]. Using more uniformly distributed datasets would improve the performance but in medical tasks it is generally costly and not feasible.

To address the class imbalance different approaches are adopted, the common ones being: Re-sampling, data augmentation, synthetic data generation, class weighting, anomaly detection techniques, ensemble methods, and the most common one in segmentation tasks: region based losses. These methods are briefly outlined here.

In resampling methods the effect of the samples of the minority classes is increased by over-sampling the minority classes or under-sampling the majority classes. For example in [21] the bicycle class which includes finely annotated segments is duplicated in the dataset. Increasing the training time, increasing the overfitting probability, causing the loss of potentially useful data in the majority classes are the drawbacks of this technique [28]. When increasing the number of samples from the minority classes, to reduce the chance of overfitting, data augmentation techniques are used. A more advanced approach would be oversampling trough synthetically creating samples from the minority classes which is proposed in [34].

Class weighting method, which is a type of cost-sensitive learning, assigns higher weights to minority classes. For example in balanced cross entropy (CE) loss the balancing weights are the inverse of class frequency [4, 17, 18, 33] this method usually leads to higher false positives and lower performance [1, 5] and it is not commonly used. This problem is recognized in [26], they argue that because some of the minority classes maybe sufficiently represented in the data and are well classified by the network, assigning higher weights to them would reduce the overall performance of the model. Instead they propose obtaining the balancing weights dynamically based on the difficulty of classification rather than the class frequency. The difficulty of each class is determined using validation accuracy of that class after each epoch. To improve the performance of balanced CE loss it is suggested in [19] that to take into account the size of each foreground apart from its class frequency. So lesions within the same class are assigned different weights depending on their size, with smaller lesions receiving higher weights.

In anomaly detection methods, based on one-class classification (OCC) techniques, the model is trained based on the majority classes and the minority classes are treated as anomalies [35]. The common strategy is that during training the input images are encoded in feature space and a decision is made for an enquiry image based on its distance in feature space or its reconstruction error in image space [35]. In medical tasks the majority class represents the normal tissue while the minority class corresponds to tumor tissue [36].

[1] University of Tehran, smhosseini741@gmail.com

Ensemble learning is another approach to handle imbalanced data. Bagging, boosting and stacking are three types of ensemble learning [37]. Bagging is when multiple models are trained on different subsets of the training data, so that some models are more focused on the minority classes. The results of different models are then combined to produce a single prediction. Boosting is when each model focuses on the misclassifications of the previous model. And stacking is the combination of multiple models with different architecture trained on the same dataset.

The most popular method to address class imbalance in segmentation tasks is the Dice loss. Dice loss is a region based loss and is derived from the Dice similarity coefficient (DSC) which measures the mismatch between the predicted and ground truth areas. Unlike the CE loss which is adopted directly from classification tasks, Dice loss is based on a geometrical metric. Existing loss functions in medical segmentation tasks are usually derived from CE and Dice loss. A review of medical segmentation losses can be found here [1].

Because Dice loss is based on the mismatch between two areas, it is inherently invariant to object size [5], and more sensitive to small errors in the predictions compared to CE loss. When there is class imbalance in the data Dice loss produces better results in segmentation of smaller objects. This property of Dice loss and also the fact that Dice similarity score is a popular network evaluation metric, have made Dice loss a widely adopted loss in medical segmentation tasks.

Sensitivity of Dice loss towards small objects can sometimes make training more difficult. As small errors due to mislabelling in ground truth, which is very common in medical tasks, can lead to large gradients. In [15] calculating Dice loss over the whole mini-batch is suggested instead of averaging individual Dice losses of each image in the mini-batch. Using the combination of CE and Dice loss, which is called compound loss, is a popular approach [15, 16, 24, 25]. In compound loss while Dice loss provides more sensitivity for smaller objects, CE loss leads to smoother gradients and better calibration [25], producing better results compared to Dice loss alone [1].

While Dice loss is effective in handling moderate class imbalance it may not perform well when we have severe class imbalance and the task involves segmenting very small objects. To address this problem modified versions of Dice loss have been proposed. In generalized Dice loss [22], similar to weighted CE losses, a balancing weight which is the squared inverse of class frequency is included in the Dice score calculation.

Dice loss is the harmonic mean of precision and recall and treats false positives and false negatives with equal importance. Missing small objects in the segmentation will cause a high false negative. To put more emphasis on false negatives [23] suggests using Tversky loss which is based on Tversky similarity index. This loss assigns a higher weight to false negatives to help with generalization and improve the performance.

Another type of imbalance in data is Difficulty imbalance. This is when the majority of pixels or voxels are easily classified and the difficult areas are in the minority. The common approach to handle difficulty imbalance is to assign higher weights to difficult pixels or voxels. The criteria for difficulty is usually based on the prediction error [2, 3] but it can also be based on prior knowledge like the position of a pixel in the image. For example, In U-net [4] higher weights are assigned to border pixels which are expected to be more difficult to classify.

Reweighting based on prediction error is mainly done through applying a modulating term [2] or resampling [3, 27]. In the first category we have the Focal CE loss [2] where the CE loss of pixels are dynamically modulated based on the prediction error to focus the training on difficult pixels. In the second category we have [27] where pixels with errors lower than a threshold are ignored. The threshold is changed based on the performance of the network. A more advance version of this method is TopK CE loss [3] where pixels are first ranked based on their prediction error and then only the top K% of pixels with highest errors are kept and the rest are ignored.

There is a correlation between difficulty imbalance and class imbalance, as the minority classes are also usually difficult to classify. But techniques that are used for class imbalance are not necessarily useful for difficulty imbalance. For example, down-weighting the majority classes, which is common in methods for class imbalance, would also down-weight the difficult areas that exist in majority classes, this would hurt the performance. And inversely if we only focus on difficult pixels because the majority of difficult pixels are most likely from the majority class the bias towards the majority class would get worse. Different methods have been proposed to address the class and difficulty imbalance at the same time.

A common strategy to handle both types of imbalance is using a compound loss: reweighted CE loss + Dice loss. Reweighted CE loss handles difficulty imbalance and Dice loss handles class imbalance. DiceFocal (Focal CE loss + Dice loss) [1] and DiceTopK (topK CE loss + Dice loss) [29] are the two popular compound losses. But the success of these methods is limited as the most sate of the art models just use CE loss + Dice loss [15, 16, 25].

The main motivation of this study is to propose a modified Dice loss which handles difficulty and class imbalance at the same time. This has been the subject of previous studies. One group of methods apply the modification at class level [38, 39, 40] and the other group at pixel or voxel level [41, 42] which is the strategy adopted in this study.

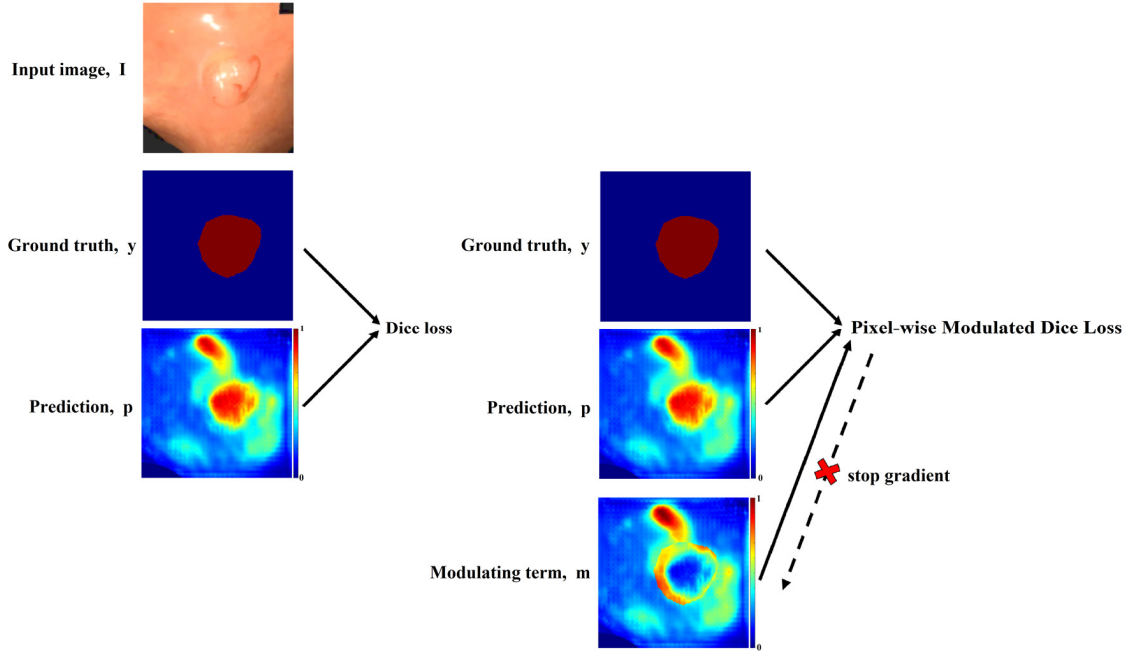

Figure 1. Pixel-wise Modulated (PM) Dice loss. In addition to the ground truth (y) and prediction ($p$) a modulating term ($m$) is also included in the calculation of the Dice loss. The modulating term is a dynamically changed hyperparameter defined as: $m_i = (|y_i - \hat{p}_i|)^\gamma$ where $\hat{p}$ is the same as $p$ but no gradient update is done through $\hat{p}$.

Logarithmic Dice loss [38] proposes replacing the Dice score of each class with its logarithmic form to increase the effect of classes with lower Dice scores. [39] Proposes the root squared version of Dice score of each class to be used. An improved version of this method with better performance is proposed in [40]. These methods would increase the effect of difficult classes with lower Dice scores on the overall Dice loss. But they modify the pixels of each class together and their performance is lower compared to methods with pixel level modification [41, 42].

The main idea of [41] and [42] is calculating the Dice loss over difficult pixels which have been chosen by a resampling strategy. In Focal Dice loss [41] true negatives are ranked based on their error values and the ones that are not in top K% are ignored. In TopK Dice loss [42] false negatives and true positives are ranked based on their error values and the ones that are not in top K% are ignored.

Different from [41, 42] the proposed method in this study uses a modulating term, as shown in Figure 1, for pixel level modification of Dice loss. This would provide more control over focusing on difficult areas. Results from experiments on three popular medical segmentation tasks show that the proposed method with a simple modification to Dice loss produces better results compared to existing methods which are designed to tackle the difficulty imbalance.

The main contributions of this work can be summarized as:

- A simple and effective modification for focusing the Dice loss on difficult areas.
- Improved segmentation accuracy, especially around boundaries, compared to other common approaches for addressing the difficulty imbalance.
- Improvements in segmentation are consistent over different types of datasets with moderate to severe imbalance. This shows that the proposed loss is versatile.
- The computational cost is minimal. And there is no need for ranking of the error values for each image, as it is required in Focal Dice loss [41], TopK CE loss and TopK Dice loss [42].
- Unlike common approaches, there is no need to combine the proposed loss with a CE based loss in a compound loss setting to improve its performance.

## 2. Method

The main motivation behind the proposed method is to take advantage of the effectiveness of Dice loss in handling the class imbalance to also handle difficulty imbalance.

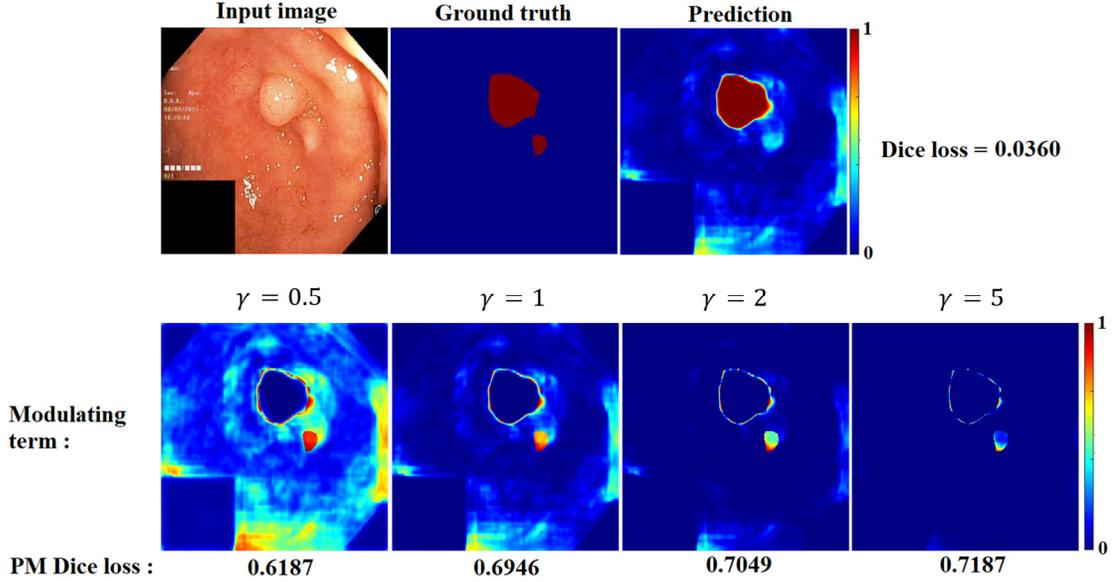

Figure 2. PM Dice loss for different focusing parameter values: $\gamma = 0.5, 1, 2, 5$. The Dice loss, which corresponds to $\gamma = 0$, is low even though the smaller polyp is almost missed by the network. In PM Dice loss the modulating term is close to zero in easy to classify areas, like the inside of the larger polyp, and close to one in difficult to classify areas, like the smaller polyp and also the boundary of the larger polyp. By increasing the focusing parameter the modulating term focuses on more difficult areas producing higher loss values. The image is from the Kvasir dataset.

The novel modification applied to Dice loss is inspired by the re-weighting strategies employed in CE based losses.

The most common CE re-weighting strategies are TopK CE loss [1, 3] and Focal CE loss [2]. Which are defined in equations 2 and 3.

$$CE\ loss = -\frac{1}{N}\sum_{c=1}^{C}\sum_{i=1}^{N} y_i^c \log(p_i^c) \quad (1)$$

$$TopK\ CE\ loss = -\frac{1}{N}\sum_{c=1}^{C}\sum_{i \in K} y_i^c \log(p_i^c) \quad (2)$$

$K$ : the set of top K% most difficult pixels

Where, $y^c$ is the ground truth label of class c, and $p^c$ is the predicted probability map for class c. N is the number of pixels, and C is the number of classes.

Easy pixels have lower errors but because they are in the majority they dominate the overall loss. TopK CE loss focuses the training on difficult pixels by removing the loss of easy pixels. The number of pixels that are kept is determined by the hyperparameter K. Lower K values mean more focus on harder pixels. But, depending on the training data there is a large variation in the number of difficult pixels, so the K value that is suitable for one image may not be able to focus on difficult pixels in another image.

To provide more flexibility Focal CE loss uses a modulating factor that determines the importance of each pixel based on its error. Less accurately predicted pixels are dynamically given a higher weight. Focal CE loss provides more control in the re-weighting process.

$$Focal\ CE\ loss = -\frac{1}{N}\sum_{c=1}^{C}\sum_{i=1}^{N} (1-p_i^c)^\gamma\ y_i^c \log(p_i^c) \quad (3)$$

With Focal CE loss easy pixels are significantly down-weighted but because of their majority they could still dominate the overall loss. This makes this loss less effective when we have a high difficulty imbalance. The second problem is that, loss of difficult pixels of a smaller class could be overwhelmed by the loss of difficult pixels from a larger class. This would exacerbate the class imbalance and reduce the performance for rare classes.

To avoid these problems we propose including a modulating term in the Dice loss to focus it on difficult pixels. The dynamic modulating term is applied at pixel level similar to the Focal CE loss. This gives the proposed method pixel level control over focusing the training on difficult to classify areas. And because of the sensitivity of Dice loss towards small objects [5], the effect of the difficult areas on the loss is less likely to be overwhelmed by the majority easy pixels.

Following [15] the variate of the Dice loss which is employed in this study is based on the Dice scores averaged over different classes, as defined in equation 4. In the denominator squared terms are used following [32].

$$Dice\ loss = 1 - \frac{1}{C}\sum_{c=1}^{C}\frac{2\sum_{i=1}^{N} y_i^c p_i^c + \epsilon}{\sum_{i=1}^{N} [(y_i^c)^2 + (p_i^c)^2] + \epsilon} \quad (4)$$

Where, $y^c$ and $p^c$ are the ground truth labels and predictions of the network for class c. N is the number of pixels, and C is the number of classes which are present in the image including the background class. $\epsilon$ is a smoothing factor to avoid division by zero.

Based on Dice loss the proposed method which we refer to as Pixel-wise Modulated Dice loss (PM Dice loss) is defined as:

$$PM\ Dice\ loss = 1 - \frac{1}{C}\sum_{c=1}^{C}\frac{2\sum_{i=1}^{N} m_i^c y_i^c p_i^c + \epsilon}{\sum_{i=1}^{N} m_i^c[(y_i^c)^2 + (p_i^c)^2] + \epsilon} \quad (5)$$

Where the modulating term for class c is: $m_i^c = (|\ y_i^c - \hat{p}_i^c\ |)^{\gamma_c}$ and $\hat{p}^c$ is the same as $p^c$ but with no gradient update. Similar to PM Dice loss we have Voxel-wise Modulated (VM) Dice loss with a 3D modulating term.

The modulating term ($m^c$) is treated as a 2D hyperparameter. No gradient is updated through $m^c$. Without this key condition the performance would drop. This is because $m^c$ is based on the prediction values, if it is not treated as a hyperparameter the loss function would no longer be similar to Dice loss. The resulted loss function cannot be considered a modified version of Dice loss, and reducing it during training would not result in predictions with better Dice scores, which serve as the evaluation metric. To avoid this, $\hat{p}^c$ and subsequently $m^c$ are treated as 2D hyperparameters. But they are not fixed, they dynamically change based on the output of the network. Also because the modulating term is a hyperparameter its computational impact on the overall gradient calculations is minimal.

$\gamma_c$ Determines the amount of focusing on difficult pixels for class c. we refer to this parameter as the focusing parameter similar to [2]. The default setting is $\gamma = 1$ where the modulating term is just the absolute value of the difference between prediction and target label. With $\gamma = 0$ the PM Dice loss becomes the original Dice loss.

When the majority of pixels are classified correctly with high confidence after applying the modulating term the numerator and denominator would be very small. To avoid getting small loss values in these cases we chose a small value for the smoothing factor, and set it to $\epsilon = 10^{-6}$.

The formula for PM Dice loss is similar to Dice loss except for the modulating term $m^c$. This term measures the distance of the prediction value and target value for each pixel. This modification to the Dice loss is much simpler compared to [41, 42] where ranking of the error values is needed for each image. The modulating term puts more emphasis on pixels with higher errors and then the Dice score is calculated. For each pixel, the more accurately it is predicted the less impact it would have on the loss.

Dice loss is more sensitive towards errors in classes with smaller areas [5] and less sensitive towards errors in classes with larger areas like the background class. This problem is addressed in PM Dice loss as the modulating term effectively reduces the area of each class to its difficult sections. Therefore the sensitivity of PM Dice loss for each class is not affected by the size of that class. As a result the loss of the background class is higher producing better precision values compared to the Dice loss.

By increasing $\gamma_c$ focus would be increased on more difficult pixels of class c. Different $\gamma_c$ values for different classes could be used to put more emphasis on certain classes. This provides a class level control over difficulty imbalance similar to [38, 39, 40] in addition to the pixel level control. This approach is adopted for the MSSEG dataset. A lower focusing parameter is assigned to the background class to prevent its loss from dominating the overall loss. More details can be found in the results section.

Figure 2 shows the effect of the modulating term with different focusing parameters on PM Dice loss. It can be seen that the original Dice loss is very low even though the smaller polyp is almost completely missed by the network. This is because the smaller polyp, which is more difficult to classify, makes up a small percentage of the total area of the two polyps. The larger polyp, which is easier to classify, is predicted by the network correctly and with high confidence leading to a low Dice loss. In PM Dice loss by focusing on difficult areas the loss values increase significantly. The modulating term is close to zero in easy to classify areas, like the inside of the larger polyp, and close to one in difficult to classify areas, like the smaller polyp and also the boundary of the larger polyp. This causes the network to focus more on difficult pixels. Figure 2 also shows the effect of γ. With higher γ values the modulating term is focused on more difficult areas and the loss is increased.

Because the modulating term is based on the predictions of the network, as the predictions get more accurate during training the modulating term also changes. This would keep the loss high and the network remains focused on the difficult areas during training.

[41] And [42] also apply a pixel level modification to the Dice loss. TopK Dice loss [42] uses a binary mask to focus the Dice loss on difficult pixels. Applying this binary mask is a resampling strategy that excludes easy pixels from Dice loss in a similar fashion to TopK CE loss. For each class the binary mask is constructed in a way so that false negatives and true positives with low errors are under-sampled. Similarly in [41] the true negatives are under-

| Loss functions | mDice | mIoU | mNSD | mPrec. | mRec. |
|---|---|---|---|---|---|
| CE | 89.56 | 83.76 | 51.13 | **92.68** | 90.30 |
| Dice | 88.76 | 82.79 | 50.84 | 91.09 | 90.33 |
| Logarithmic Dice [38] | 88.89 | 82.86 | 49.49 | 90.97 | 90.75 |
| Focal Dice [40] | 88.19 | 81.46 | 46.27 | 88.78 | 91.24 |
| Focal Dice [41] | 89.12 | 83.40 | 51.22 | 91.17 | 90.74 |
| CE + Dice | 89.94 | 84.20 | 52.34 | 92.58 | 90.67 |
| Focal CE + Dice | 89.85 | 84.29 | 52.15 | 92.20 | 91.11 |
| TopK CE + Dice | 90.11 | 84.47 | 52.56 | 92.23 | 91.17 |
| TopK Dice [42] | 89.94 | 84.11 | 52.76 | 91.62 | 91.42 |
| CE + TopK Dice [42] | 89.62 | 83.99 | 52.80 | 91.71 | 91.24 |
| TopK CE + TopK Dice [42] | 89.39 | 83.53 | 52.75 | 91.17 | 90.97 |
| PM Dice (our) | **90.61** | **85.37** | **54.90** | 92.61 | 91.57 |
| CE + PM Dice (our) | 90.45 | 85.20 | 54.21 | 92.64 | 91.40 |
| Focal CE + PM Dice (our) | 90.58 | 85.26 | 54.14 | 92.25 | **91.68** |

Table 1. Results of the Kvasir dataset. The best and second best results are shown in red and blue respectively.

| Loss functions | mDice | mIoU | mNSD | mPrec. | mRec. | RV | MYO | LV |
|---|---|---|---|---|---|---|---|---|
| CE | 91.60 | 84.88 | 87.05 | 92.36 | 91.19 | 90.61 | 88.90 | 95.29 |
| Dice | 90.02 | 82.73 | 86.02 | 88.72 | 92.35 | 86.06 | 89.20 | 94.80 |
| Logarithmic Dice [38] | 89.60 | 82.12 | 85.23 | 88.21 | 92.11 | 85.28 | 88.92 | 94.61 |
| Focal Dice [40] | 88.72 | 80.81 | 83.38 | 87.26 | 91.47 | 83.54 | 88.28 | 94.34 |
| Focal Dice [41] | 90.34 | 83.21 | 86.41 | 88.83 | 92.87 | 86.32 | 89.53 | 95.16 |
| CE + Dice | 91.41 | 84.63 | 86.99 | 91.62 | 91.65 | 90.10 | 89.26 | 94.87 |
| Focal CE + Dice | 91.61 | 84.93 | 87.05 | 91.78 | 91.85 | 90.30 | 89.22 | 95.30 |
| TopK CE + Dice | 91.75 | 85.12 | 87.34 | 92.16 | 91.70 | 90.54 | 89.30 | 95.40 |
| TopK Dice [42] | 90.34 | 83.22 | 86.38 | 88.77 | **92.98** | 86.30 | **89.69** | 95.04 |
| CE + TopK Dice [42] | 91.75 | 85.15 | 87.38 | 91.49 | 92.42 | 90.49 | 89.40 | 95.35 |
| TopK CE + TopK Dice [42] | 91.84 | 85.26 | 87.53 | 91.78 | 92.26 | 90.69 | 89.45 | 95.37 |
| PM Dice (our) | **91.88** | **85.35** | **87.92** | 92.34 | 91.82 | 90.54 | 89.43 | **95.67** |
| CE + PM Dice (our) | 91.68 | 85.03 | 87.61 | **92.48** | 91.31 | 90.35 | 89.11 | 95.57 |
| Focal CE + PM Dice (our) | 91.84 | 85.28 | 87.81 | **92.48** | 91.61 | 90.58 | 89.34 | 95.60 |

Table 2. Results of the ACDC multi-organ dataset. The mean dice score of 3 organs are also reported. The best and second best results are shown in red and blue respectively.

sampled. These methods rely on one or two fixed hyperparameters to focus the loss on difficult pixels. This is not a very effective strategy considering the large variation in difficulty levels between images and during training. Different from [41] and [42] we use a dynamically changed 2D modulating factor which is more effective, as it can be seen in the results section.

## 3. Experiments

To evaluate the performance of the proposed method multiple experiments on three popular medical segmentation datasets are conducted. The training conditions, like the network architecture, training data, augmentation techniques, and learning rate, are kept the same in different experiments with the only different factor being the loss function.

### 3.1. Datasets

The datasets are chosen in a way to include different types of medical segmentation tasks. The tasks are binary and multi-class segmentation with class and difficulty imbalance ranging from mild to severe.

**Kvasir-SEG dataset [43]:** this is a polyp segmentation task containing 1,000 polyp RGB images and their corresponding binary ground truth labels. Following [46] the images are split into 90% for training and 10% for testing. The resolution of images ranges from $332 \times 482$ to $1920 \times 1072$. All images are resized to $320 \times 320$ and patches of size $256 \times 256$ are randomly cropped during training.

**Automated Cardiac Diagnosis (ACDC) dataset [44]:** this dataset contains cardiac 3D MRI scans of 150 patients. The classes in this multi-class segmentation task are: left ventricles (LV), right ventricles (RV), and the myocardium (MYO). Following [12], out of 100 scans of the training set, 70 and 20 scans are randomly chosen for training and testing. The input slices are resized to $256 \times 256$ and normalized to [0, 1] before training and testing.

| Loss functions | mDice | mIoU | mNSD | mPrec. | mRec. |
|---|---|---|---|---|---|
| CE | 65.46 | 49.06 | 62.38 | 72.13 | 62.07 |
| Dice | 66.41 | 50.07 | 63.67 | 71.67 | 63.82 |
| Logarithmic Dice [38] | 65.94 | 49.57 | 62.96 | 72.36 | 62.47 |
| Focal Dice [40] | 63.48 | 46.88 | 60.04 | 67.85 | 61.85 |
| Focal Dice [41] | 66.80 | 50.57 | 64.30 | 71.29 | 64.90 |
| CE + Dice | 67.23 | 50.99 | 64.57 | 72.49 | 64.71 |
| Focal CE + Dice | 67.92 | 51.77 | 65.54 | 74.87 | 63.67 |
| TopK CE + Dice | 67.16 | 50.89 | 64.62 | 73.24 | 63.62 |
| TopK Dice [42] | 68.36 | 52.44 | 66.23 | 70.66 | **68.14** |
| CE + TopK Dice [42] | 67.36 | 51.26 | 64.97 | 70.78 | 66.73 |
| TopK CE + TopK Dice [42] | 68.44 | 52.51 | 66.25 | 72.56 | 66.39 |
| PM Dice (our) | **69.07** | **53.12** | **66.93** | 75.15 | 64.92 |
| CE + PM Dice (our) | 68.71 | 52.80 | 66.73 | **76.46** | 63.60 |
| Focal CE + PM Dice (our) | 68.72 | 52.78 | 66.33 | 74.30 | 65.40 |

Table 3. Results of the MSSEG dataset. The best and second best results are shown in red and blue respectively.

**MSSEG-2 dataset [45]:** this dataset consists of 15 brain scans, and the task is the binary segmentation of multiple sclerosis (MS) lesions. The ground truth labels are the consensus between 4 different expert annotations. 9 scans were chosen randomly for training and the rest for testing. All 5 available modalities are used as the input of the network. The input slices are resized to 256 × 256 and normalized to [0, 1] before training and testing.

### 3.2. Implementation

The popular U-Net [5] is chosen to study the performance of the proposed loss function. This is following [1] where a survey of a large set of different loss functions in medical segmentation tasks is done using U-Net. For Kvasir and ACDC datasets a U-Net structure with the backbone of ResNet34 [6] is implemented. The output of the network at scale 1/4 is resized to the original size by bilinear up-sampling during training and inference [13, 14]. For MSSEG dataset a simple U-Net [5] with 4 max pooling layers is utilized. Weight decay is set to $10^{-4}$ and the Adam optimizer is used. The learning rate is reduced according to a polynomial learning rate policy (initial learning rate $\times (1 - \text{iter.}/\max\_\text{iter.})^{0.9}$) with the initial learning rate of $5 \times 10^{-4}$ for the U-Net with ResNet34 backbone and $10^{-4}$ for the simple U-Net. Results are the average of 3 independent runs. For PM Dice loss and other losses based on Dice loss the smoothing factor is set to $\epsilon = 10^{-6}$. Data augmentation techniques include random flipping, rotation, zooming, and elastic deformation.

In PM Dice loss for Kvasir and ACDC datasets the default value of $\gamma = 1$ is used for all classes, which produces the best results with the least computational cost. For MSSEG dataset where the imbalance is more severe we use a higher focusing parameter for the foreground compared to the background, the values are set to: $\gamma_{FG} = 2$, $\gamma_{BG} = 1$. The effect of choosing different focusing parameter values is investigated in the results section, Figures 3 and 4.

The evaluation metrics include the mean Dice score (mDice), mean Intersection over Union (mIoU), mean precision (mPrec.) and mean recall (mRec.) of the foreground classes. In ACDC dataset, the average Dice score of each organ is also reported.

For evaluating the quality of segmentation boundaries, following [1], Normalized Surface Distance (NSD) [30] is utilized. NSD, defined in equation 6, is a combination of boundary based and region based metrics [31]. Like the Dice score NSD is bounded between 0 and 1. NSD = 0 means there is no overlap between the boundary of the ground truth and the boundary of the predicted region, and NSD = 1 means there is a complete overlap between the two boundaries.

$$NSD = \frac{\sum [(S_y \cap B_p^{(\tau)}) + (S_p \cap B_y^{(\tau)})]}{\sum [S_y + S_p]} \quad (6)$$

Where, $S_y$ and $S_p$ are the borders of ground truth and predicted regions respectively. Around these borders, based on the threshold value τ, the boundary regions $B_y^{(\tau)}$ and $B_p^{(\tau)}$ are defined.

### 3.3. Results

The qualitative comparison between different methods are presented in this section. As it can be seen in tables 1 to 3 the proposed method produces better results compared to other methods. A visual comparison of the results of different loss functions is shown in figure 5.

A few patterns can be observed from the results which we examine here. When comparing the Dice and CE loss, we see that generally, Dice loss produces higher recall values and the CE loss produces higher precision values.

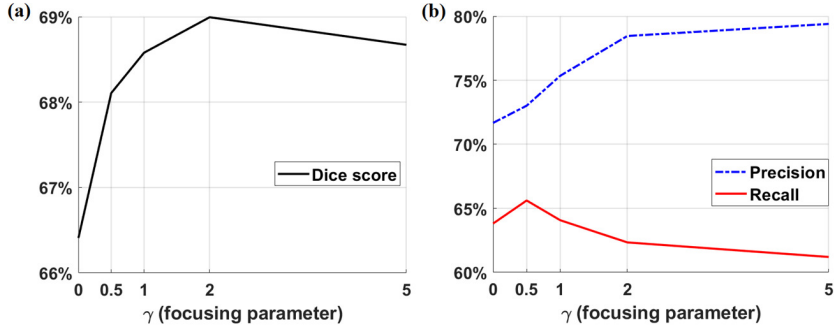

Figure 3. The effect of changing the focusing parameter ($\gamma$) on mean Dice score, precision and recall values in MSSEG dataset. $\gamma$ is the same for the foreground and the background class.

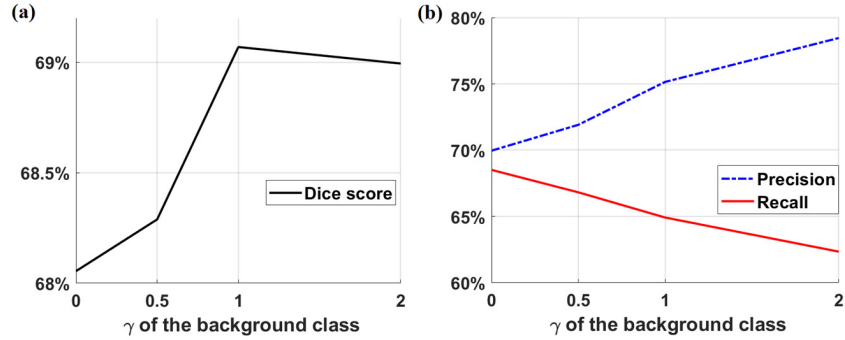

Figure 4. The effect of changing the focusing parameter of the background class ($\gamma_{BG}$) on mean Dice score, precision and recall values in MSSEG dataset. The focusing parameter of the foreground class is set to: $\gamma_{FG} = 2$.

This is because Dice loss is dominated by the loss of classes with smaller areas, and CE loss is dominated by the loss of the background class. The combination of these two losses (compound loss = CE + Dice) provides a balance between the two losses and usually improves the results as it can also be seen in [1].

Focal CE and TopK CE focus more on difficult pixels and because the majority of pixels are from the background class these losses increase the sensitivity of CE towards the background class. That is why Focal CE + Dice and TopK CE + Dice usually produce higher precision values compared to CE + Dice. Focal Dice loss [41] focuses on false positives and true negatives with low confidence, as a result this loss produces higher recall values compared to Dice loss. TopK Dice loss [42] by only considering the top K% of difficult pixels, reduces the area of each class. This increases the sensitivity towards errors in the classes with smaller areas compared to classes with larger areas like the background class. This increased sensitivity leads to higher recall values compared to Dice loss. An effective combination is TopK CE + TopK Dice loss which generally produces better results compared to CE + Dice loss.

The proposed PM Dice loss outperforms other loss functions which are designed to handle difficulty imbalance in all three datasets. Unlike Dice loss precision values produced by PM Dice loss is also high and therefore combining it with a CE based loss is not required to improve its performance. High precision values of PM Dice loss is due to the fact that focusing on difficult areas would increase the effect of the background class on the overall loss. This is in contrast with Dice loss which is more sensitive towards classes with smaller areas [5] and less sensitives towards the larger background class. PM Dice loss produces better results compared to Dice loss and also when it replaces Dice loss in a compound loss setting.

In all three datasets the proposed PM Dice loss produces the highest NSD values, which means more accurate predictions around boundaries. This is because the proposed loss focuses on difficult areas and the boundary areas are usually the most difficult parts of the image to classify especially in medical tasks where objects usually don't have clear borders. This shows that the proposed loss is more effective in focusing on difficult areas, like the boundary areas. Improvements in boundary segmentation also happens when the PM Dice loss is combined with a CE based loss.

By adjusting the focusing parameter the performance of the PM Dice loss can sometimes be improved compared to the default value of $\gamma = 1$. This is especially the case in the MSSEG dataset, where the imbalance in the data is more severe. The effect of changing the focusing parameter ($\gamma$) on the performance in the MSSEG dataset is investigated in Figure 3. Similar to [2] the values chosen for the focusing parameter are 0, 0.5, 1, 2 and 5. These values are applied to both foreground and background class. With $\gamma = 0.5$ higher precision and recall values are produced compared to $\gamma = 0$ (original Dice loss). By further increasing $\gamma$ precision keeps increasing but recall decreases, and the best Dice score is achieved by $\gamma = 2$. In datasets where there is a high imbalance between the foreground and the background class, using higher focusing parameter values would cause the loss of the background class to dominate the overall loss, leading to high precision and low recall values. This is because in these datasets the network correctly classifies almost all of the background pixels with very high confidence. This results in a modulating term that would focus the loss on a very small percentage of background pixels, which are very difficult to classify, leading to higher loss values for the background class.

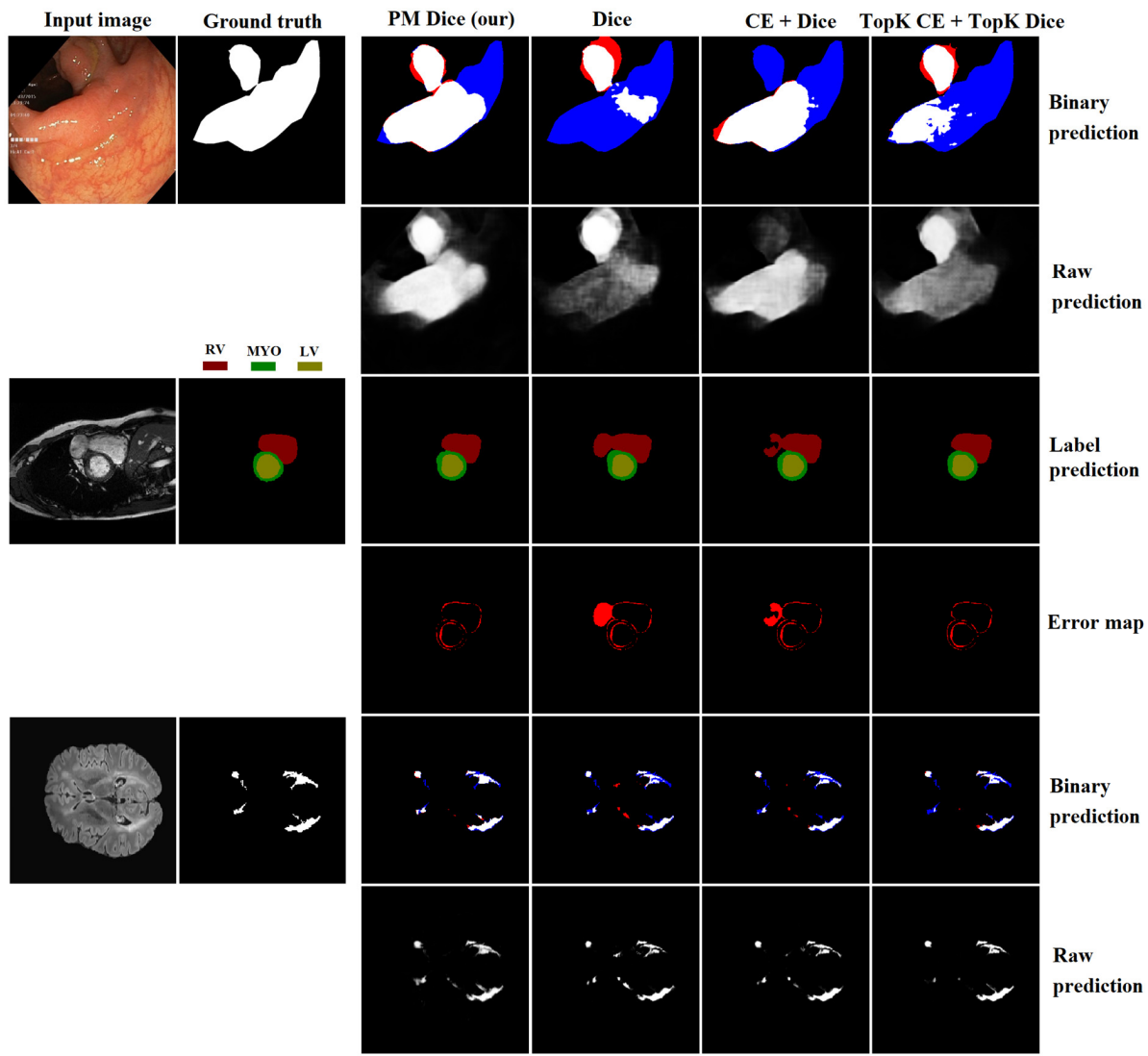

Figure 5. Visual comparison of results from Kvasir, ACDC and MSSEG segmentation tasks. In binary segmentation tasks false positives and false negatives are marked by red and blue respectively.

By using different focusing parameters for the foreground and the background class we can reduce the impact of the loss of the background class on the overall loss. Figure 4 shows the effect of using different $\gamma$ values for different classes in MSSEG dataset. The focusing parameter of the foreground class is set to: $\gamma_{FG} = 2$ and for the background class we have: $\gamma_{BG} = 0, 0.5, 1, 2$. With lower values of $\gamma_{BG}$ the impact of the background class is reduced, leading to lower precision and higher recall values. The highest Dice score is achieved by: $\gamma_{FG} = 2$, $\gamma_{BG} = 1$. The results reported in Table 3 are obtained using these focusing parameters.

## 4. Conclusion

In this study, a simple and effective modification to the Dice loss is proposed to enable it to handle difficulty imbalance as well as class imbalance. A modulating term reduces the effect of easily classified pixels on the overall loss focusing the training on difficult areas. Experiments on three widely used datasets show that the proposed loss improves segmentation accuracy, especially around boundaries, outperforming other common loss functions which are designed to tackle difficulty imbalance. The improvements are achieved with minimal computational cost, and with no need for a CE based loss in a compound loss approach.